# Hierarchy of Hofstadter states and replica quantum Hall ferromagnetism in graphene superlattices

G. L. Yu[1], R. V. Gorbachev[2], J. S. Tu[2], A. V. Kretinin[2], Y. Cao[2], R. Jalil[2], F. Withers[2], L. A. Ponomarenko[1,3], B. A. Piot[4], M. Potemski[4], D. C. Elias[1,5], X. Chen[3], K. Watanabe[6], T. Taniguchi[6], I. V. Grigorieva[1] , K. S. Novoselov[1], V. I. Fal'ko[3], A. K. Geim[1,2*], A. Mishchenko[1*]

[1]School of Physics & Astronomy, University of Manchester, Manchester M13 9PL, UK

[2]Centre for Mesoscience & Nanotechnology, University of Manchester, Manchester M13 9PL, UK

[3]Physics Department, Lancaster University, LA1 4YB, UK

[4]Laboratoire National des Champs Magnétiques Intenses, CNRS-UJF-UPS-INSA, F-38042 Grenoble, France

[5]Departamento de Física, Universidade Federal de Minas Gerais, 30123-970, Belo Horizonte, Brazil

[6]National Institute for Materials Science, 1-1 Namiki, Tsukuba, 305-0044 Japan

*In graphene placed on hexagonal boron nitride, replicas of the original Dirac spectrum appear near edges of superlattice minibands. More such replicas develop in high magnetic fields, and their quantization gives rise to a fractal pattern of Landau levels, referred to as the Hofstadter butterfly. Some evidence for the butterfly has recently been reported by using transport measurements. Here we employ capacitance spectroscopy to probe directly the density of states and energy gaps in graphene superlattices. Without magnetic field, replica spectra are seen as pronounced minima in the density of states surrounded by van Hove singularities. The Hofstadter butterfly shows up in magnetocapacitance clearer than in transport measurements and, near one flux quantum per superlattice unit cell, we observe Landau fan diagrams related to quantization of Dirac replicas in a reduced magnetic field. Electron-electron interaction strongly modifies the superlattice spectrum. In particular, we find that graphene's quantum Hall ferromagnetism, due to lifted spin and valley degeneracies, exhibits a reverse Stoner transition at commensurable fluxes and that Landau levels of Dirac replicas support their own ferromagnetic states.*

When graphene is placed on top of atomically flat hexagonal boron nitride (hBN) and their crystallographic axes are carefully aligned, graphene's electron transport properties become strongly modified by a hexagonal periodic potential induced by the hBN substrate[1-6]. Replicas of the main Dirac spectrum appear[7-12] at the edges of superlattice Brillouin zones (SBZ) and, for the lowest SBZs, the second-generation Dirac cones can be reached using electric field doping[4-6]. Because the superlattice period, $\lambda$, for aligned graphene-hBN structures is relatively large ($\approx$15 nm), magnetic fields $B \sim$10 T are sufficient to provide a magnetic flux $\Phi$ of about one flux quantum $\phi_0$ per area $A = \sqrt{3}\lambda^2/2$ of the superlattice unit cell. The commensurability between $\lambda$ and the magnetic length $l_B$ gives rise to a fractal energy spectrum, the Hofstadter butterfly[4-6,13-19]. An informative way to understand its structure is to consider the butterfly as a collection of Landau levels (LLs) that originate from numerous mini-replicas of the original spectrum, which appear at all rational flux values $\Phi = \phi_0 (p/q)$ where $p$ and $q$ are integer[4,12]. At these fluxes, the electronic spectrum can be described[12-15] in terms of Zak's minibands[14] for an extended superlattice with a unit cell $q$ times larger than the original one. In graphene, Zak's minibands are expected to be gapped cones (third-generation Dirac fermions)[12]. Away from the rational flux values, these Dirac replicas experience Landau quantization in an effective field $B_{eff} = B - B_{p/q}$ where $B_{p/q} = \phi_0(p/q)/A$.

In this work, we have employed capacitance measurements to examine the electronic spectrum of graphene superlattices and its evolution into the Hofstadter butterfly. In zero $B$, pronounced minima in the electronic density of states (DoS) are observed not only for graphene's neutral state but also at high electron and hole doping. The latter minima signify Dirac replicas near edges of the first SBZ. The replicas occupy a spectral width of $\sim$50 meV indicating strong superlattice modulation. Temperature ($T$) dependence of the DoS suggests that the second-generation Dirac cones are singly and triply degenerate for graphene's valence and conduction bands, respectively. In quantizing $B$, in addition to the classic fan diagram for graphene, we observe many new cyclotron gaps that fan out from finite values of $B$. They are attributed to the formation of high-field replica Dirac cones[4] and their Landau quantization in $B_{eff}$. The local fans are particularly well developed near $\Phi = \phi_0$ and $\nu = 0$, $\pm 1$, $\pm 2$ where $\nu = n\phi_0/B$ is the filling factor and $n$ the carrier density. The Hofstadter minigaps in this regime cannot be explained by orbital quantization only. They do not follow the expected dependence on the LL index and are described by the Coulomb energy scale $E_C = e^2/\varepsilon l_B^*$ where $l_B^*$ is the magnetic length in $B_{eff} = \pm |B - B_{1/1}|$ and $\varepsilon$ the effective dielectric constant[20-22]. We also observe that the SU(4) quantum Hall ferromagnetism (QHFM), characteristic of nonaligned devices[20-23], experiences strong suppression at commensurable fluxes. The $\nu = \pm 1$ gaps disappear near $\Phi = \phi_0$ whereas the ferromagnetic states at $|\nu| = 3$, 4 and 5 exhibit a reentrant transition at $\Phi = \phi_0/2$.

**DoS for second-generation Dirac fermions**

Figure 1a shows our capacitor devices. Graphene is placed on top of hBN (50-100 nm thick) and encapsulated with the second hBN crystal of thickness $d$. A gold electrode is then evaporated on top. The whole structure is fabricated on a quartz substrate to minimize parasitic capacitances. The devices are similar to those studied previously[24] but a critical step is added: crystallographic alignment of graphene and one of the encapsulating hBN crystals with a precision of $\sim 1°$ by using procedures of ref. 4. Seven capacitors with areas $S$ ranging from 50 to 350 $\mu m^2$ and $d \approx$10 to 40 nm have been studied. Depending on accuracy of our alignment, densities $n$ at which the first SBZ becomes fully filled are found between $\approx$3 and 6 $\times 10^{12}$ cm$^{-2}$. All the studied capacitors exhibit little residual doping (Fig. 1a), and their charge carrier mobilities vary from $\approx$50,000 to 120,000 cm$^2$V$^{-1}$s$^{-1}$, as found using Hall bar devices fabricated in parallel with the capacitors. The differential capacitance $C$ was measured by an on-chip bridge made following the recipe of ref. 6. Note that Hofstadter states were not observed in the previous capacitance measurements[6].

A typical behavior of $C$ in zero $B$ as a function of bias $V_b$ applied between graphene and the Au electrodes is shown in Fig. 1a. A sharp minimum near zero $V_b$ corresponds to the main neutrality point where the DOS tends to zero and remains finite only due to charge inhomogeneity[24]. There are additional minima at large electric-field doping (Fig. 1a). Following the earlier analysis[4-6], the features can be attributed to second-generation Dirac cones (inset in Fig. 1a). It is possible to translate $C$ directly into the DoS or, equivalently, the quantum capacitance $C_Q = Se^2 DoS$ where $DoS$ is the density of states and $e$ the electron charge. To this end, we write[24] $C = (1/C_Q + 1/C_G)^{-1} + C_p$ where $C_G$ and $C_p$ are the geometric and parasitic (parallel) capacitances, respectively. $C_G$ is defined by the known values of $d$ and $S$ and can be measured independently from the periodicity of magnetocapacitance oscillations[24]. This leaves $C_p$ as the only fitting parameter to determine $C_Q$. In our devices, $C_p$ is ~1 fF (<1% of $C$) and, in most cases, no fitting is required. To present $DoS$ as a function of the Fermi energy $E$ rather than $V_b$, we first integrate $C(V_b)$ curves over $V_b$, which yields the induced $n$, and then subtract the electrostatic voltage drop from applied bias[24]: $E = eV_b - e^2 n/C_G$ (again, no fitting parameters). Examples of this conversion of $C(V_b)$ into $DoS(E)$ are shown in Fig. 1b. In this presentation, the spectral changes induced by the superlattice potential become clear and more pronounced. Instead of the standard linear behavior ($DoS \propto |E|$) seen for non-aligned graphene-on-hBN capacitors[24,25], deep minima in the DoS appear for $|E| > 0.15$ eV in both valence and conduction bands of graphene. The minima are surrounded by equally pronounced maxima. The behavior is manifestation of the Dirac replicas formed at the edges of the first SBZ and terminated by van Hove singularities (left inset of Fig. 1a). This agrees with the earlier results[3] obtained by scanning tunneling spectroscopy where qualitatively similar but smeared $DoS$ curves were observed. The spectral width occupied by the second-generation Dirac fermions (distance between van Hove singularities) is ≈25 and ≈75 meV for positive and negative $E$, respectively. The nonmonotonic behavior at $|E| > 0.2$ eV is reproducible for different devices and indicates contributions coming from further SBZ.

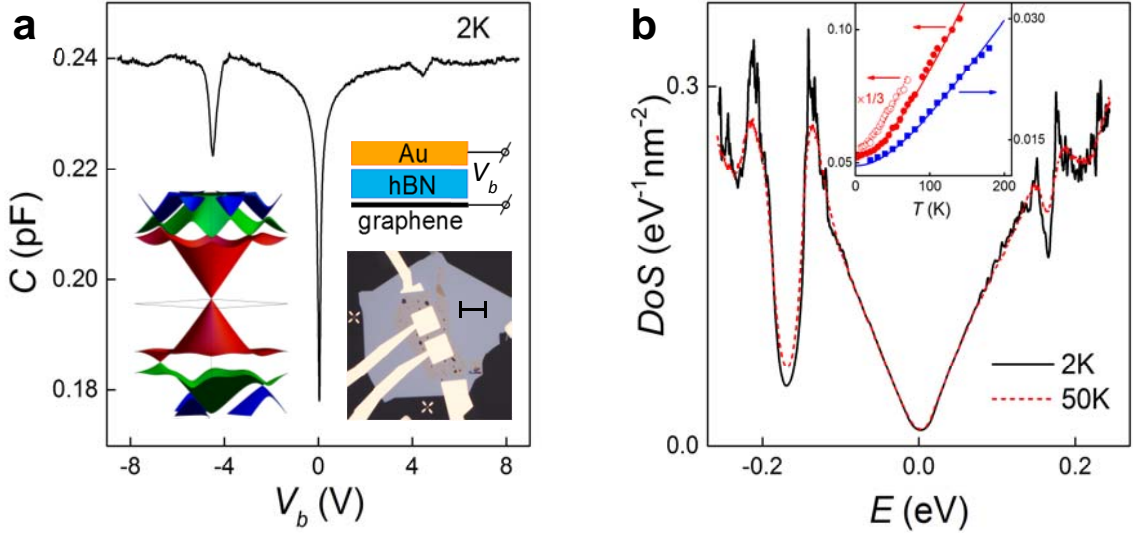

FIG. 1. **Capacitance spectroscopy of graphene superlattices. a** – Typical $C(V_b)$ in zero magnetic field. Positive and negative $V_b$ correspond to electron and hole doping, respectively. The range of applied

$V_b$ is limited by dielectric breakdown of hBN (<0.5 V/nm). For the shown capacitor, $S$ =308 μm$^2$ and $d$ =35 nm. Right insets: Schematics of our devices and their optical image. The two square regions are Au gates, and the top and bottom leads are electrical contacts to graphene. Scale bar: 20 μm. Left inset: One of the theoretically proposed scenarios for the low-$E$ band structure of graphene on hBN[11,12]. The first, second and third SBZ are shown in red, green and blue, respectively. **b** – DoS for the device in **a**. The inset shows $T$ dependence of the DoS in the three minima (blue symbols, main DP; solid red, hole-side DP; open red, electron-side DP). Solid curves: theory fits. Note that the smearing of the DOS by scattering and inhomogeneity is less than 50 K.

Taking into account graphene's spin and valley degeneracy, we write $DoS = 8\pi g_S |E|/h^2 v_F^2$ where $h$ is the Planck constant and $g_S$ the additional degeneracy of secondary Dirac spectra (inset of Fig. 1a). The observed linear behavior at $|E|$ <0.1 eV in Fig. 1b yields $v_F^0$ =0.98±0.04×10$^6$ ms$^{-1}$, in agreement with literature values. The slopes $DoS \propto |E|$ extrapolate to zero DoS with no indication of a sizeable gap (>5 meV) open by the hBN potential[4,6,26]. Further information about secondary Dirac fermions can be obtained by analyzing $T$ dependences of the DoS (Fig. 1b). At a Dirac point (DP), the DoS is expected to increase linearly with $T$ as[27] $DoS(T) = 8\pi \ln(4) g_S T/h^2 v_F^2$. For the main spectrum, $g_S/v_F^2$ is known and no fitting parameter is needed to describe the observed behavior at high $T$ (inset of Fig. 1b). The saturation at low $T$ is due to charge inhomogeneity, which we model by the Gaussian distribution of $n$ with a standard deviation $\delta n$. The blue curve in the inset of Fig. 1b is for $\delta n \approx 1.3 \times 10^{10}$ cm$^{-2}$, in agreement with the smearing of $DoS$ as a function of $E$ at low $T$. For hole- and electron- side replicas, their DoS at high $T$ increases ≈5 and 15 times, respectively, quicker than that at the main minimum. By taking into account that $v_F$ is expected[3-12] to be ≈$v_F^0$/2 and assuming that $\delta n$ does not significantly change with $V_b$, the data in the inset of Fig. 1b yield $g_S$ =1 and $v_F \approx 0.45 \times 10^6$ ms$^{-1}$ for the hole-side Dirac cones and $g_S$ =3 and $v_F \approx 0.4 \times 10^6$ ms$^{-1}$ for the electron one (red curves). These values are also consistent with the observed $DoS(E)$ curves. The found $g_S$ mean that there are one and three Dirac replicas for negative and positive $E$, respectively, in agreement with the spectral reconstruction scenario shown in the inset of Fig. 1a. Note that charge inhomogeneity does not allow us to eliminate the possibility of small gaps at the main and secondary DPs[6,11].

**Capacitance spectroscopy of Hofstadter butterfly**

Figure 2 shows examples of $C(V_b)$ in the regime of quantizing $B$. Landau quantization results in numerous sharp minima that develop in $B$ >1 T, so that at 3 T one can see all the cyclotron gaps for $\nu$ up to 38 (Fig. 2a). Many-body gaps at $\nu$ = 0 and ±1 (due to the lifted spin-valley degeneracy) open up at ≈1.5 and 4 T, respectively. In low $B$, we can employ the same approach as described above to convert $C(V_b)$ into $DoS(E)$. Examples are given in Figs 2d and S1. One can see sharp peaks in the DoS which correspond to metallic LLs (incompressible states with $C \approx C_G$) separated by wide regions of a low DoS (cyclotron gaps). The conversion procedure cannot be applied automatically to strongly quantizing $B$ because in the quantum Hall regime the bulk becomes increasingly isolated from electric contacts, and central areas no longer contribute to the measured signal. This leads to excessively deep minima in $C$ (effectively, smaller $S$) and large systematic errors in determining $E$ (Supplementary Information; section 1). We can increase $T$ to suppress the insulating state but because of different gap sizes this has to be done individually for each gap and each range of $B$ (Fig. S1). Therefore, our high-$B$ data are mostly presented in the original format, as $C(V_b)$.

In addition to the cyclotron gaps described above, we find numerous magnetocapacitance minima that are not observed in similar but non-aligned graphene-on-hBN devices[24]. The minima become very pronounced in high $B$ (Figs 2b-c), and their evolution leads to a complex pattern seen in Fig. 3a. For clarity, the pattern is reproduced schematically in Fig. 3b. The observed fan diagram is consistent with those reported previously[4-6] but rather unexpectedly the DoS measurements reveal more Hofstadter states than could be seen in the transport experiments.

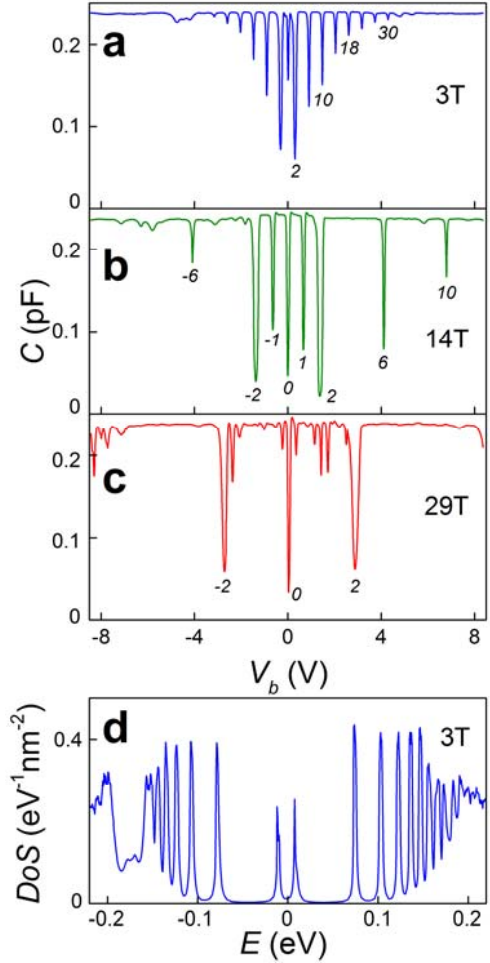

**FIG. 2. Magnetocapacitance oscillations and quantized DoS. a-c** – Typical $C(V_b)$ in various $B$ at 2 K. The numbers mark $\nu$ for the corresponding minima. **d** – DoS found from **a**.

Colors in Fig. 3b mark four types of gaps. First, there are the expected single-particle gaps originating from the main DP (black lines). Second, the blue lines indicate many-body gaps due to QHFM. Third, there are minima in the DoS which fan out from hole- and electron- side DPs (green). Intuitively, they could be attributed to LLs for second-generation Dirac fermions. However, this would be wrong because it is easy to estimate that the cyclotron energy becomes comparable to the spectral width of the Dirac replicas already in $B$ <1T. Accordingly, all the LLs should bunch at van Hove singularities (Fig. 1b) before even being resolved in our measurements. Comparison with theory[4,12] suggests that the green gaps represent Hofstadter minibands in the regime of strong superlattice modulation where any distinction between LLs originating from the main and superlattice spectra is lost. Fourth, our fan diagrams show minima that cannot be traced back to either the main or secondary DPs (red). We attribute them to LLs originating from replica Dirac spectra that are quantized in $B_{eff} = \pm|B - B_{1/q}|$ with $q$ = 1, 2, … This is seen most clearly around $\Phi = \phi_0$ where minigaps spray up and down from $\nu$ = 0, ±1, ±2 and -6 (Fig. 3). This is the case of relatively weak modulation such that individual LLs split into superlattice minibands without intermixing[16-18]. As discussed above, the superlattice band structure at exactly $\Phi = \phi_0$ consists of replica Dirac cones[4,12] and, for positive and negative $B_{eff}$, these cones experience quantization which leads to local fan diagrams. Note that the red and green gaps in Fig. 3b have essentially the same (i.e. superlattice) origin and we have distinguished between them above only to facilitate the discussion.

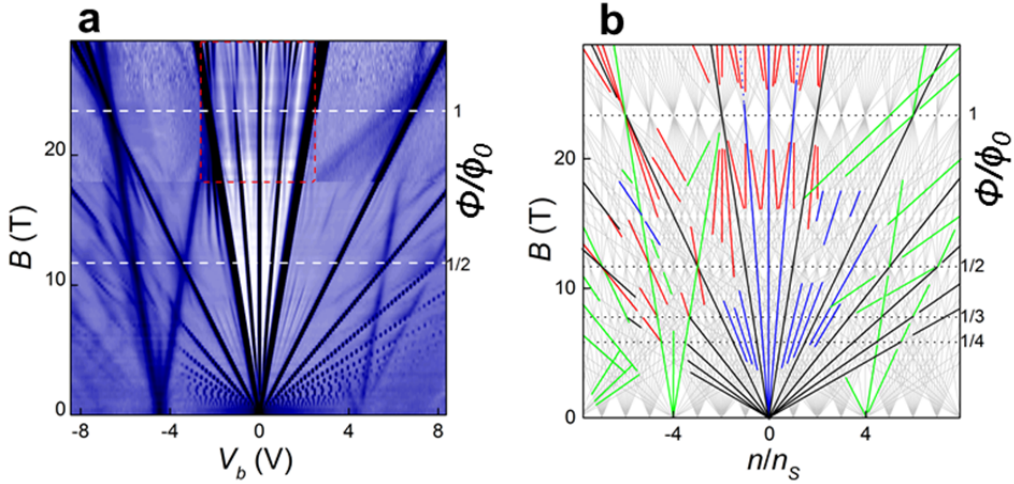

**FIG. 3. Hofstadter butterfly in graphene superlattices. a** – Fan diagram $C(V_b,B)$. Gaps appear as dark stripes. Scale: navy-to-white, 0.9 to 1.02 $C_G$; black, $C$ <0.9 $C_G$. Such zooming does not distinguish between deep and shallow gaps that both appear dark. The data are for the device in Fig. 2. Our other superlattice capacitors show similar behavior but weakest minima may disappear due to charge inhomogeneity (Supplementary Information; section 2). The measurements were done in a superconducting magnet (up to 18T) and in an electromagnet of Grenoble's High Magnetic Field Laboratory (up to 29T). The presented diagram is a combination of the two sets. Data within the red dashed rectangle (|$\nu$| <2) were accumulated over several days to improve resolution. **b** – DoS minima from **a** shown schematically. The grey lines are the Wannier grid[28] that shows allowed gaps including those with lifted spin and valley degeneracies. 4×$n_S$ is the carrier density at which the first SBZ is filled. $\Phi = B \times A$ for the right axis is calculated using the known value of $n_S$ =1/$A$ which yields $A$ directly[4-6].

**Interaction effects in Hofstadter spectrum**

The simultaneous occurrence of QHFM and Hofstadter states suggests a possibility of their interplay. Several observations show that this is indeed the case. Most striking among them is the suppression of QHFM at commensurable fluxes $\Phi/\phi_0$ =1 and 1/2 (Fig. 4). One can notice this effect already in Fig. 3 where the QHFM gaps at $\nu$ =3, ±4, ±5 disappear around $\Phi = \phi_0/2$ (also, see Fig. S4). The suppression of QHFM is further elucidated in Figs 4b-c. After the onset of QHFM in a few T, the main many-body gaps gradually grow with increasing $B$ but then shrink approaching $\Phi = \phi_0$. The QHFM gaps at $\nu$ =±1 vanish around $\Phi = \phi_0$, and there are weaker minima near $\Phi = \phi_0/2$ (clearly developed for $\nu$ =-1; see Fig. 4c). This behavior has not been observed in graphene-on-hBN without alignment[24] (Supplementary Information).

We attribute the suppression of QHFM to a reverse Stoner transition. According to numerical modelling over a broad range of moiré parameters[4,12], the zero LL remains relatively narrow for $B$ <0.7 $B_{1/1}$ but then exponentially broadens into a sizeable band of width $D$ (Supplementary Information; Fig. S5). As the spin-valley ferromagnetism specifically requires narrow LLs, the superlattice broadening of a LL should result in a tendency of the system to return into the normal metal state or, at least, reduce the associated QHFM gap by $D$. In moderate $B$, the QHFM gap at $\nu=0$ is accurately described by $E_C = e^2/\varepsilon l_B \propto \sqrt{B}$ (inset of Fig. 4c). By extrapolating this behavior to high $B$ (black curve in Fig. 4c), we can estimate $D$ of the zero LL from the relative drop in the measured gap, $\Delta = E_c - D$. This yields $D \approx 5$ meV at $\Phi \approx \phi_0$. The complete suppression of the QHFM at $\nu = \pm 1$ is attributed to low energy costs of creating skyrmion-like spin/valley textures[20,21,29-32], which in combination with sizeable $D$ close the gaps (see Supplementary Information; Fig. S8).

**FIG. 4. Interactions in Hofstadter minibands. a** – Zoom into the high-$B$ region of Fig. 3a. Diameters of the yellow circles indicate relative sizes for superlattice minigaps at 29 T. They are marked as ($\nu, \nu_L$) where the filling factor $\nu$ indicates the association with the $\nu$-th gap in the main spectrum, and

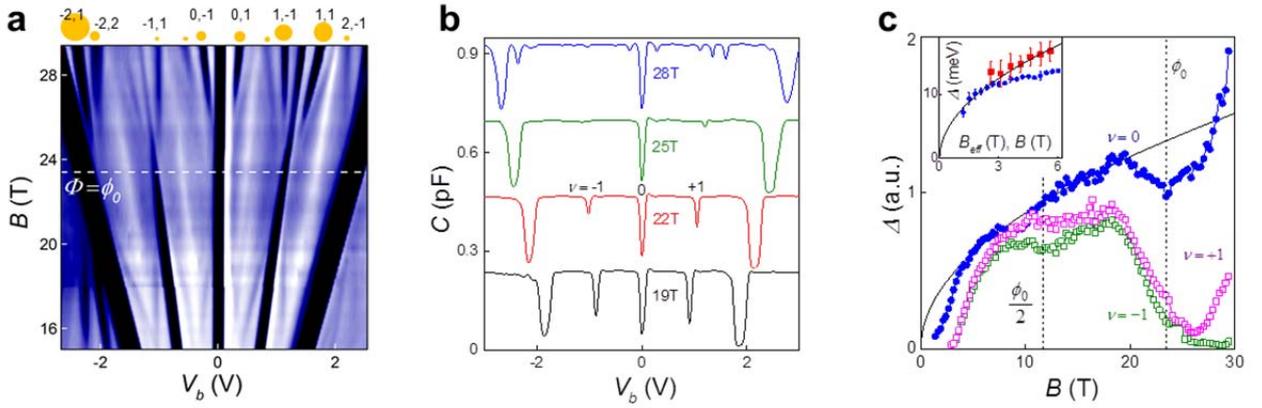

$\nu_L$ is the local filling factor. The latter is determined from the slopes $B(n)$ in the fan diagram in Fig. 3b, which yield a value of ($\nu_L - \nu$) (see Supplementary Information). **b** – Cross-sections from **a** to illustrate the closure of the $\nu = \pm 1$ gaps around 25 T. **c** – Evolution of the main spin-valley gaps with $B$. The gaps are obtained by integration of $C(V_b)$ curves. Inset: Largest minigap (-2,1) as a function of $B_{eff}$ (red symbols) and the $\nu = 0$ gap as a function of $B$ (blue). The gaps are found from $T$ dependences of capacitance minima at fixed $B$ (ref. 24). Both $\Delta$ trail the Coulomb energy scale $E_C$ with $\varepsilon = 8$ for encapsulated graphene[24] (solid curve). $T = 2$ K for all the plots.

Figure 4 provides a closer look at incompressible superlattice states ($\nu, \nu_L$) near $\Phi = \phi_0$. One may naively attribute them to cyclotron gaps for third-generation Dirac fermions in fields $B_{eff} = B - B_{1/1}$. However, states ($\pm 2, \nu_L$) fanning out from $\nu = \pm 2$ contain pronounced gaps at odd $\nu_L$ (Fig. 4a). Because the $\nu = \pm 2$ QHE states are neither spin nor valley polarized, one cannot explain these odd-integer gaps simply by superlattice effects, without invoking spontaneous lifting of the degeneracies. Furthermore, the inset in Fig. 4c shows one of the odd-integer minigaps (-2,1) as a function of $B_{eff}$ and compares it with the QHFM gap at $\nu = 0$ plotted as a function of $B$ (the latter's behavior agrees with the earlier report[24]). Both gaps $\Delta$ exhibit absolute values close to $E_C = e^2/\varepsilon l_B \propto \sqrt{B^*}$ where $l_B = (h/2\pi eB^*)^{1/2}$ and $B^*$ is given by either $B$ or $B_{eff}$ (Fig. 4c). This agreement points at the same origin for the two gaps. The above reasoning also applies to the fan diagram (0, $\nu_L$) which contains odd $\nu_L$. The odd-integer minigaps (0,±1) exhibit the same dependence $\Delta \propto \sqrt{B_{eff}}$ but smaller absolute values.

Our analysis of the effect of electron-electron interaction on LLs of third-generation Dirac fermions is given in Supplementary Information (section 4) and provides qualitative understanding of the entire hierarchy of the observed minigaps. For example, the replica quantum Hall ferromagnetism allows us not only to understand the presence of large gaps at (-2,1) and (-2,2) but also the virtual absence of similarly indexed states at (-2,-1) and (-2,-2) (see Fig. 4a).

**Conclusions**

Capacitance spectroscopy is a powerful tool for investigating graphene superlattices and reveals directly numerous Hofstadter minigaps. Outside a strongly-insulating quantum Hall regime, capacitance data can accurately be converted into the density of states. Our observations elucidate the concept that Hofstadter minibands can be viewed as Landau quantization of replica Dirac cones. In high $B$, the Hofstadter butterfly is strongly influenced by Coulomb interaction. The most spectacular result is the collapse of quantum Hall ferromagnetism at commensurable fluxes $\Phi = \phi_0$ and $\phi_0/2$, which is attributed to a reverse Stoner transition and is likely to be related to the formation of skyrmion-like textures. Quantitative description for the observed interaction effects remains to be developed but their strength and variety indicate a large playground to study many-body phenomena in Dirac systems, which is relatively easily accessible by magnetocapacitance measurements.

*Acknowledgements* - This work was supported by the European Research Council, the Royal Society, Graphene Flagship and EuroMagNET II (EU Contract 228043).

# Supplementary Information

## #1 Conversion of magnetocapacitance into energy spectra

As described in the main text, the procedure of converting $C(V_b)$ into $DoS(E)$ generally fails in quantizing $B$ because graphene electrodes become highly insulating in the quantum Hall effect regime. This results in the measured minima in $C$ to become excessively deep. Even small inaccuracies in their depth, which do not visibly distort $C(V_b)$ curves, may lead to notable errors in $E$ because of the required integration (see the main text). In this section, we illustrate how such errors look like and discuss how they can be avoided.

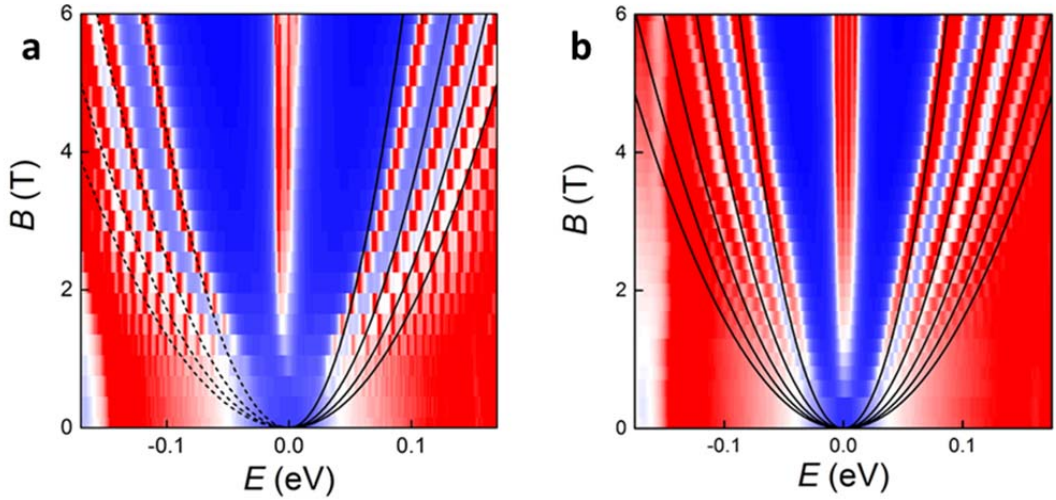

**Fig. S1. Quantized spectrum of graphene from magnetocapacitance.** The color maps are made of DoS curves such as those shown in Fig. 2d of the main text. Blue-to-red scale: $DoS$ changes from 0 to $C_G/Se^2$. The rectangular pattern is due to a finite step size in $B$: the curves are taken at every 0.3 and 0.2 T for left and right panels, respectively. **a** – $T$ =2 K. The LL positions found from capacitance cannot be described accurately by eq. 1 (black solid curves for $E$ >0). However, the errors come mostly from the overestimated size of the $\nu$ =2 gap. If the position of the first LL ($N$ =1) is taken as the base line, the $N$ =2 and 3 levels follow eq. 1 (dashed curves for $E$ <0). **b** – At 30 K, the insulating state at $\nu$ =±2 is sufficiently suppressed, and all the resolved LLs follow the correct $B$ dependence given by eq. 1 (black curves).

Figure S1 shows an energy spectrum of graphene obtained by converting $C(V_b)$ curves measured in different $B$ into a $DoS(E,B)$ map by naively following the procedures described in the main text and ref. 24. Maxima in $DoS$ in Fig. S1 (shown in red) indicate LLs. Qualitatively, they follow the $\sqrt{B}$ dependence, characteristic for graphene, and the absolute energy scale is correct. For quantitative analysis, we use the known expression for LL energies in graphene

$$E_N = v_F \cdot (2e\hbar BN)^{1/2} \qquad (1)$$

where $v_F \approx 10^6$ m/s is the Fermi velocity and $N$ the LL index. The expected $E_N$ are plotted in Fig. S1 as solid black curves. One can see that the positions of LLs in Fig. S1a agree with the expected values only for $B$ up to ≈2T. In higher $B$, the first LL goes significantly higher in $E$ than expected, and the deviations from its correct position grow with $B$. These errors (~20%) originate from the fact that the $\nu = \pm 2$ state becomes increasingly insulating. After we know that the position of the first LL is inferred wrongly, we can check whether our measurements of the other cyclotron gaps at $\nu = \pm 6$ and $\pm 10$ are similarly affected by their insulating behavior. It turns out to be not the case. This is shown in Fig. S1a by the dashed curves, which are calculated taking the position of the $N = -1$ LL as the base level and adding the corresponding energy gaps $\Delta$ between higher LLs according to eq. 1. The agreement shows that only the $\nu = \pm 2$ state is insulating enough over the plotted range of $B$ to distort our magnetocapacitance measurements while the smaller cyclotron gaps are measured correctly.

For accurate measurements of the $\nu = \pm 2$ gap, we need to increase $T$ which suppresses the insulating behavior in graphene's bulk (Fig. S1b). One can see that the $N = \pm 1$ LLs have now moved into their correct positions whereas the other levels remain at their previous positions with respect to $N = \pm 1$. Unfortunately, LLs also broaden with increasing $T$ which makes it difficult to measure small gaps accurately at higher $T$. Therefore, the discussed conversion of magnetocapacitance into the DoS has to be implemented for each gap using different intervals of $T$ and $B$.

To avoid the conversion errors described above, one can estimate that resistivity of graphene should be significantly less than $1/fC \sim 10^{10}$ Ohm where $f$ = 1kHz is the frequency of our measurements. As a rule of thumb, the conversion procedure should not be used from the moment when a small out-of-phase capacitance signal can be detected, even though its contribution to the in-phase signal may seem negligible.

#2 Further examples of Hofstadter butterflies

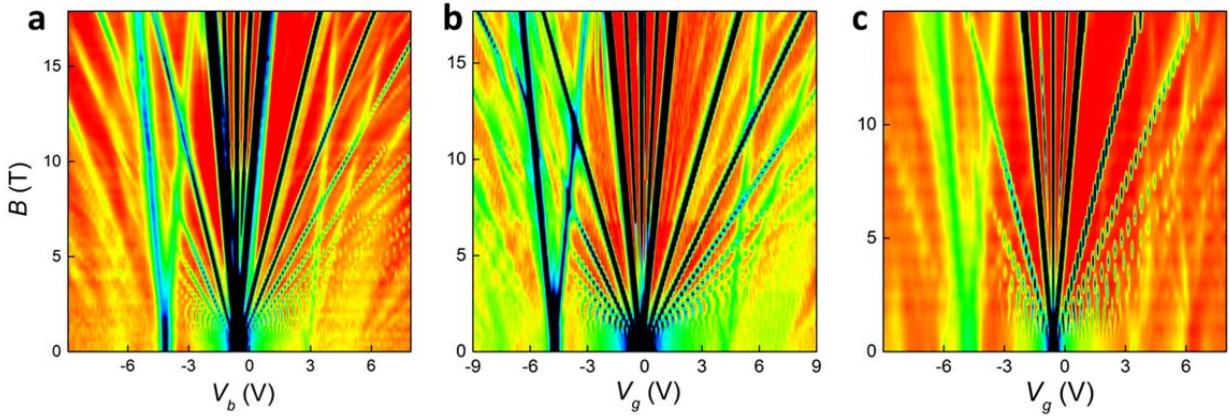

**Fig. S2. Hofstadter butterflies for three other aligned graphene-on-hBN devices.** The dotted pattern (most pronounced in low $B$) appears due to a finite step in $B$, which is used to accumulate data for such maps by sweeping $V_b$. The measurements were carried out at 2 K in a superconducting magnet allowing $B$ up to 18 T. **a** – $C(V_b,B)$ for a capacitor with $d \approx 28$ nm. Scale: blue-to-green-to-red, 0.95 to 0.98 to 1.01 $C_G$; black, $C$ <0.95 $C_G$. One may notice weak lines running in parallel to the strongest capacitance minima (e.g., to the right of the $\nu = +6$ gap). This behavior suggests that a small part of the device has a neutrality point shifted by $\approx 0.2$ V. **b** – Device with $d \approx 35$ nm. Scale: blue-to-green-to-red, 0.97 to 0.99 to 1.01 $C_G$; black, $C$ <0.97 $C_G$. **c** – $d \approx 37$ nm. Scale: blue-to-green-to-red, 0.93 to 0.97 to 1.0 $C_G$; black, $C$ <0.93 $C_G$.

Figure S2 shows magnetocapacitance maps measured for several other graphene-on-hBN devices with accurately aligned crystallographic axes of graphene and hBN. For all these devices the condition $\Phi = \phi_0/2$ is reached close to 12 T. The observed behavior quantitatively agrees with that in Fig. 3 of the main text over the same range of $B$. Another slightly misaligned device with $B_{1/1} \approx 36$ T exhibited a rather similar map that was stretched in $B$ accordingly (not shown). The main difference between different devices is that some shallow states disappear, presumably due to higher charge inhomogeneity. For example, Hofstadter states (-2,1), (-2,2) and (-2,3) are well resolved in $B \sim 15$ T (i.e. for negative $B_{eff} = B - B_{1/1}$) for the capacitor in Fig. 3 of the main text. One can also see these states for the device in Fig. S2b. Only one state [probably, (-2,2)] remains resolved in Fig. S2a and no such Hofstadter minigaps can be seen for the device in Fig. S2c.

#### #3 Quantum Hall ferromagnetism in aligned versus nonaligned devices

The suppression of QHFM in high $B$ could only be observed for graphene superlattices (that is, in aligned graphene-on-hBN devices). To illustrate this fact, Figs S3-S4 show the DoS behavior observed for a capacitor in which graphene and hBN crystallographic axes were deliberately misaligned by a large angle (~5°). The $C(V_b,B)$ map in Fig. S3a reveals the standard behavior expected for graphene in the absence of superlattice effects (also, see ref. 24). The many-body gaps at $\nu=0$ and $\pm1$ are clearly resolved in $B$ starting from a few T, which indicates graphene's electronic quality similar to that for the device in Fig. 3 of the main text. All the gaps including those due to QHFM increase monotonically with increasing $B$, in contrast to the behavior for the aligned capacitors in the main text. To emphasize this fact, Fig. S3b plots cross-sections of the observed map over the same range of $B$ as used in Fig. 4b of the main text.

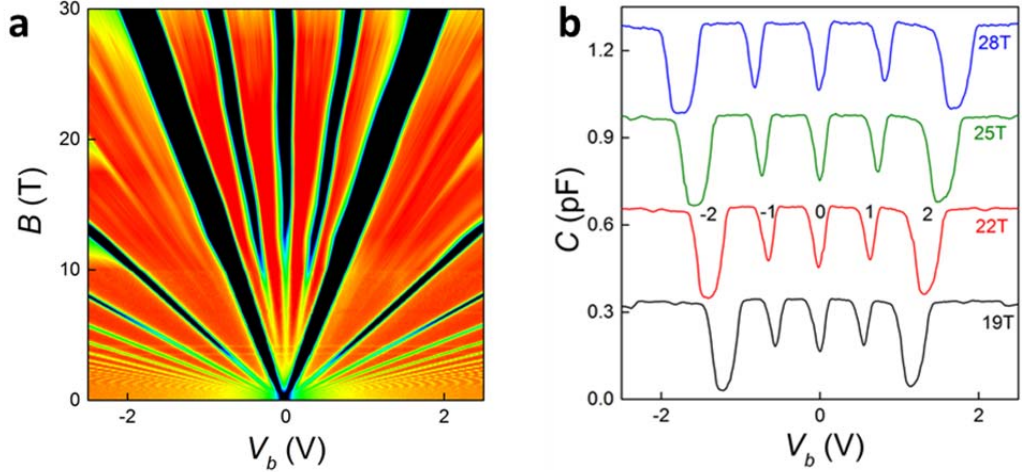

**Fig. S3. Conventional quantum Hall ferromagnetism in nonaligned graphene-on-hBN device. a –** Full map showing the development of many-body gaps at $\nu=0$ and $\pm1$. **b –** No gap closures occur in high $B$, in contrast to the behavior in Fig. 4 of the main text. Device's parameters: $S=275$ μm$^2$; $d=23$ nm; 2 K. Scale: blue-to-green-to-red, 0.76 to 0.91 to 1.04 $C_G$; black, $C<0.76\ C_G$.

The suppression of quantum Hall ferromagnetism at commensurable fluxes is further illustrated in Fig. S4. Here we show the depth $\Delta C$ of capacitance minima for $\nu=0$ and $\pm1$ (Fig. S4a). $\Delta C$ correlates with gap's size: the larger the gap, the deeper the minimum. One can see that the QHFM minima in Fig. 4a evolve very differently for aligned (solid symbols) and non-aligned (open) capacitors. For the case of graphene superlattices, the $\nu=-1$ minimum practically disappears at $\Phi=\phi_0$, in agreement with the results in Fig. 4 of the main text. Also, a small dip is seen at $\Phi=\phi_0/2$. Similar behavior is observed for the $\nu=+1$ minimum (not shown). The $\nu=0$ minimum exhibits a partial suppression in $B$

>20 T. In contrast, the same minima for non-aligned capacitors evolve monotonically within only small irregular fluctuations along the curves, which are due to lower accuracy used for non-aligned graphene devices to save 30-T magnet time.

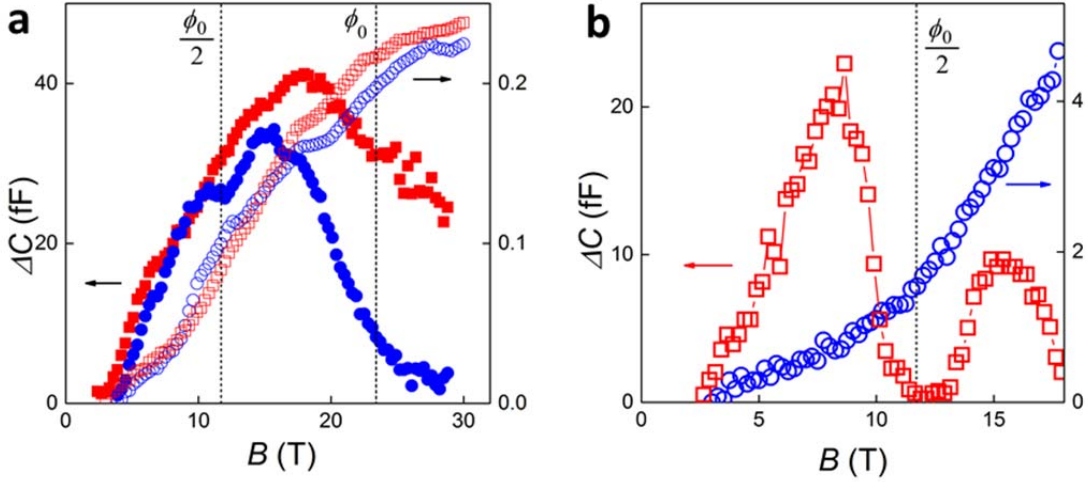

**Fig. S4. Quantum Hall ferromagnetism in graphene superlattices and graphene. a** – Depth of capacitance minima at $\nu=0$ (red) and -1 (blue) as a function of magnetic field. Solid symbols – aligned graphene on hBN (the same device as shown in Fig. S2b); open symbols – non-aligned device (same as in Fig. S3). **b** – Depth of the $\nu=+4$ minimum for aligned (red squares) and non-aligned (blue circles) devices.

Finally, we revisit the suppression of QHFM states at the $\nu=\pm3, \pm4, \pm5$ near $\Phi=\phi_0/2$, which is clearly seen in Fig. 3 of the main text and briefly discussed in there. Figure S4b elucidates this behavior by plotting the depth of the $\nu=+4$ minimum as a function of $B$. The minimum disappears at $\Phi=\phi_0/2$, then reappears again. In contrast to this reentrant behavior, non-aligned devices show a gradual increase in the depth of all their spin-valley resolved minima, including the one plotted in Fig. S4b.

### #4 Quantum Hall ferromagnetism in Hofstadter minibands

In this section, we analyze the hierarchy of the observed incompressible states (Fig. 4 of the main text). As pointed out in the main text, states ($\nu_L, \nu$) with odd-integer $\nu_L$ are not allowed in the single-particle picture and, to explain their appearance, one has to take into account the interplay between the formation of Hofstadter minibands and electron-electron interaction. The latter is known to result in spontaneously polarized electronic states which are usually referred to as QHFM states and exhibit odd-integer filling [S1].

Figure S5 shows a characteristic example of a numerically calculated spectrum for Dirac electrons in graphene superlattices in the presence of magnetic field (more examples for various choices of a moiré pattern can be found in Refs. 4,12,19). In low $B$, the low-energy part of the spectrum in Fig. S5 exhibits clearly identifiable LLs. The $N=0$ LL marks zero $E$ and is separated from the other levels by the largest cyclotron gap, $v_F \cdot (2e\hbar B)^{1/2}$. This LL broadens into a fractal band that reaches the width $D$ at $B = B_{1/1}$. The band remains exponentially narrow is $B < 0.7 B_{1/1}$ having the width $\sim u e^{-\pi\phi_0/\sqrt{3}\Phi}$ where $u$ is the amplitude of moiré modulation[11]. Below we refer to this band as central. At $B_{1/1}$, the central band is separated from the other bands in the fractal spectrum by gaps $\delta \sim [v_F \cdot (2e\hbar B)^{1/2} - D]$ on both valence and conduction sides of the spectrum in Fig. S5.

A generic feature of graphene-on-hBN superlattices is that their spectrum near $B = B_{1/1}$ resembles the quantized spectrum of gapped Dirac fermions[4,12] with the gap $\delta$ positioned between the edges of the central and neighboring bands (that is, $\delta$ is aligned with the $\nu = \pm 2$ cyclotron gaps). This is illustrated in Fig. S5 that plots the energy dispersions for two lowest magnetic minibands within the valence band at exactly $B_{1/1}$. Their top and bottom can be approximated by $E = \pm\sqrt{\delta^2/4 + v_*^2 p^2}$ where $v_*$ is the Fermi velocity for this third-generation Dirac spectrum[12]. Moreover, for $\Phi$ close to $\phi_0$, the fractal minibands can be identified as LLs of third-generation Dirac fermions in the effective field $B_{eff}$ = $B - B_{1/1}$. These LLs exhibit energy $E_N = \pm\sqrt{\delta^2/4 + 2v_*^2 \hbar e |B_{eff}| N}$.

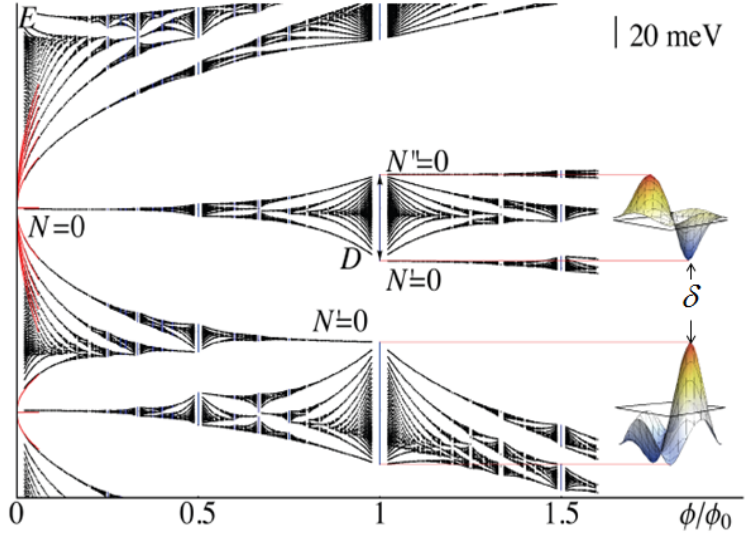

**Fig. S5. Hofstadter spectrum for zero and adjacent Landau levels.** Magnetic minibands calculated for a graphene-on-hBN superlattice using $vbu_0$ =34meV, $vbu_1$ =-67meV and $vbu_3$ =-58meV (see the Refs. 11,12 for the parameters' definition). Red curves to the left part of the figure show analytically calculated LLs for primary and secondary Dirac fermions[4,12]. The insets to the right show the energy dispersion at $\Phi = \phi_0$. To make the third-generation spectra look less complicated, we use here the rhombic rather than hexagonal shape for the first Brillouin zone.

Particularly important for our analysis below is that the spectrum features nearly dispersionless levels $N'=0$ and $N''=0$, which appear for both positive and negative $B_{eff}$ and are separated from the rest of the relatively dense spectrum by large 'cyclotron' gaps (Fig. S5). Therefore, in the following we focus on these well-separated 'zero' LLs and, for simplicity, consider the rest of the minibands as a continuous band of states (Figs. S6-S8). Note that the spectrum in Fig. S5 is obtained in the single-particle approximation, and each of the LLs is spin and valley degenerate. To account for the degeneracy, Figs. S6-S8 use colors to indicate four different spin and valley states.

It is well established [S1] that spin and valley degeneracies can be lifted by Coulomb repulsion of electrons, which leads to spontaneous polarization of the electronic system (ferromagnetism). QHFM develops in two dimensional systems at odd-integer $\nu$ [S1] where electrons occupy half the states in the highest spin-degenerate LL. For spin and valley degenerate LLs in the main graphene spectrum, this leads to the formation of spin and valley polarized states at the filling factors $\nu = \pm 1$ whereas at $\nu = 0$ an antiferromagnetic state is formed with opposite polarization in valleys $K$ and $K'$ because of stronger intra-valley interaction with respect to the inter-valley one[20-23,29-32]. The antiferromagnetic state leads to the energy gain per electron [S2-S4], $E_C(B) \sim e^2/\varepsilon l_B$. This energy also determines the gap in the electron addition spectrum [S5,S6]. On the other hand, in the fully polarized ferromagnetic states at $\nu = \pm 1$, the formation of long-range spin textures (skyrmions) reduces the addition energy to [S7-S9]

$$E_{sk}(B) = \alpha E_C(B), \ 1/2 < \alpha < 1.$$

Parameter $\alpha$ depends on a value of the single-particle Zeeman splitting, $E_Z \ll E_C$, and is related to the skyrmion radius. For $E_Z \to 0$, the skyrmion size becomes infinite and $\alpha = 1/2$ whereas, for large $E_Z$, skyrmions shrink in size and $\alpha \to 1$. Also, according to Ref. S4, no skyrmion formation occurs in the antiferromagnetic state $\nu = 0$, due to valley dependence (inter- *vs* intra-) of electron-electron interactions. The latter excludes the involvement of the valley degree of freedom in the formation of skyrmion-like textures at $\nu=\pm 1$.

Below, we consider how all these many-body effects combine with characteristic features of the fractal single-particle spectrum in Fig. S5. This allows us to explain the hierarchy of incompressible states in each of the observed sequences ($\pm 2, \nu_L$), ($\pm 1, \nu_L$) and ($0, \nu_L$) described in the main text as 'replica' ferromagnetism for the *N'=0* and *N''=0* LLs of third-generation Dirac fermions.

### A. ($\pm 2, \nu_L$) sequence

**(-2,0).** At $\nu$ =-2, the magnetic minibands shown in Fig. S5 are filled up to the largest gap in the spectrum, $\delta_{-2,0} \sim [v_F \cdot (2e\hbar B)^{1/2} - D]$, just below the central band emerging from the *N=0* LL. The fact that the central band is empty is reflected by notation (-2,0). In this case, the electrons are not polarized, and the resulting spectrum is schematically shown in Fig. S6 (left).

**(-2,1).** Adding electrons to the (-2,0) state at $B > B_{1/1}$ leads to a partial occupancy of the *N'=0* LL of third-generation Dirac fermions, which is expected to result in a ferromagnetic instability similar to that for the *N=0* LL of primary Dirac fermions at low *B*. State (-2,1) corresponds to the unit filling factor of the *N'=0* LL so that

$$\nu_L = (n_e - n_{\nu=-2})\phi_0/(B - B_{1/1}) = 1.$$

By the same reason as for the case of $\nu$ =-1 in low *B*, electrons in the *N'=0* LL can reduce their Coulomb energy by forming a fully spin-valley polarized state. The gap in this state, caused by exchange interactions, is

$$\delta_{-2,1} \approx E_C(B_{eff}) = \sqrt{\frac{\pi}{2}} e^2/\varepsilon l_{B_{eff}}$$

and determined by the magnetic length $l_B$ set by the effective magnetic field $B_{eff}$. Importantly, energy $\delta_{-2,1}$ is not expected to be reduced by the formation of skyrmion textures because the Zeeman splitting is determined by the total magnetic field ~25 T, which should be sufficient to suppress skyrmions. The resulting gapped spectrum and the occupancy of levels are illustrated in the right panel in Fig. S6.

One should also take into account that, at $B < B_{1/1}$, the *N'=0* LL for gapped Dirac fermions behaves very peculiar and appears at the upper edge of the adjacent Hofstadter band (Fig. S5). Therefore, as shown in Fig. S6 (right), for negative $B_{eff}$ the condition $\nu_L = (n_e - n_{\nu=-2})\phi_0/(B - B_{1/1}) = 1$ corresponds to a unit filling factor for holes (negative $n_e - n_{\nu=-2}$) and results in the same size of the gap, $\delta_{-2,1}$.

**(-2,2).** Similarly, we interpret the observed incompressible state (-2,2) as the fully spin polarized *N'=0* LL with both valleys occupied. Therefore, we expect $\delta_{-2,2} \sim \delta_{-2,1}$.

**(-2,-|$\nu_L$|).** With reference to the spectrum shown in Fig. S5, we point out that much smaller gaps should be expected in the case of the (-2,$\nu_L$) states with negative $\nu_L$. This is because their occupancy involves the |$N'$|=1 LL of third-generation Dirac fermions (Fig. S5). In contrast to the $N'$=0 LL, the considered sequence (-2,-|$\nu_L$|) is separated from the dense spectrum below by smaller gaps, so that the LL mixing and the increased screening [increasing $\varepsilon$ in $E_C(B_{eff})$] are expected to suppress the exchange gaps.

The above expectations for sequence (-2,$\nu_L$) agree with the observed hierarchy of the incompressible states in Fig. 4. Because of the electron-hole symmetry, the above arguments equally apply to the symmetric sequence (+2,$\nu_L$).

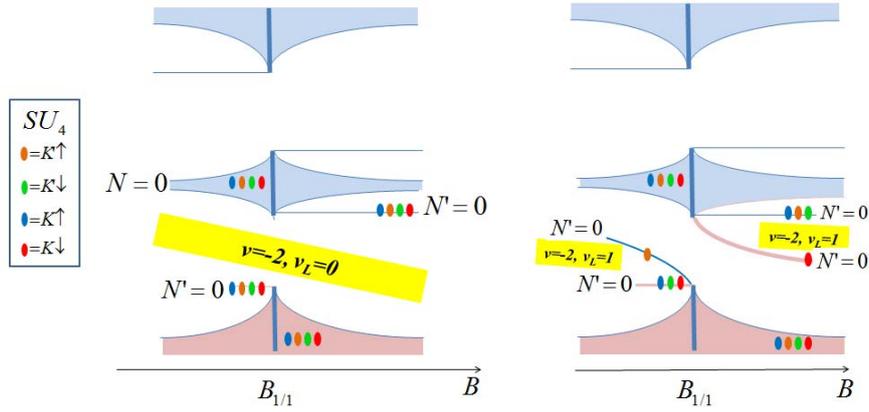

**Fig. S6. Quantum Hall ferromagnetism at the zero LL of third-generation Dirac fermions for the (±2,$\nu_L$) sequence.** Incompressible states associated with spin-valley polarized states (-2,$\nu_L$=-2,-1,0,1,2) around $B_{1/1}$. In each panel, superlattice-broadened LLs of the main graphene spectrum are shown schematically, following the calculated spectrum in Fig. S5. The central magnetic band is the $N$ =0 LL; the other bands are adjacent LLs ($N$ =±1). The filled and empty states are shown in pink and blue, respectively. $N'$ =0 marks the zero LL for the third-generation Dirac replica; the vertical axis indicates energy.

### B. (0,$\nu_L$) sequence

**(0,0).** Figure S7 (left) sketches the energy diagram for $\nu$ =0, which in terms of occupancy of third-generation LLs is equivalent to (0,0). This is a neutral state of graphene such that electrons fill half the 4-fold degenerate $N$=0 LL. It has been argued[23] that exchange interaction leads in this case to quantum Hall antiferromagnetism with opposite spin polarizations for valleys $K$ and $K'$ and, accordingly, generates an exchange gap $E_C(B)$ between the states with $K\uparrow$ and $K'\downarrow$, as illustrated in Fig. S7. Broadening of the central band at $B$ =$B_{1/1}$ is expected to result in reduction of the actual incompressibility gap in the system such that $\delta_{0,0} \approx E_C(B) - D$, in agreement with the experiment (Fig. 4c of the main text).

**(0,±1).** At positive $B_{eff}$, by adding electrons ($n_e$ >0) to the blue band in Fig. 7 (right) one fills the $N'$=0 LL of third-generation Dirac fermions. For the filling factor $\nu_L = n_e\phi_0/(B-B_{1/1})$ =1, we expect again a fully spin polarized ferromagnetic state with gap $\delta_{0,1} \sim E_C(B_{eff})$. The same argument can be applied to adding holes ($n_e$<0) to the pink band, which yields $\delta_{0,-1} \sim \delta_{0,1}$.

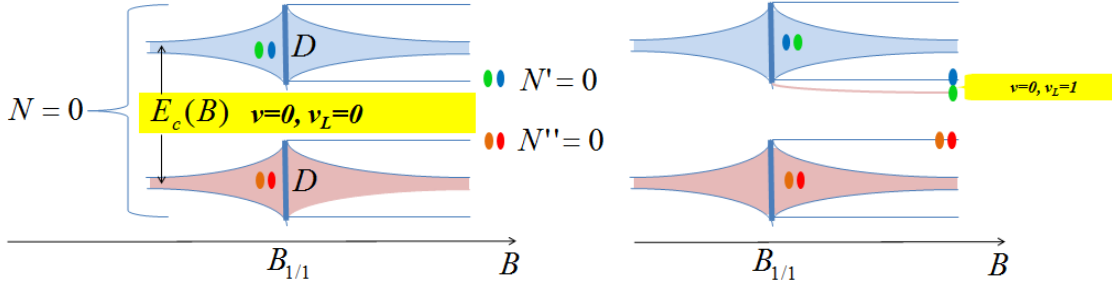

**Fig. S7. (0, $v_L$) sequence of QHFM states.** Incompressible states (0,-1) and (0,0) around $B_{1/1}$. Left panel: Reduction of the activation energy gap from its bare value, $E_c$, to $E_c$ - $D$ at $B_{1/1}$. Right panel: Effect of exchange splitting within the valley-polarized $N'$=0 Landau level. For brevity, we consider only the central band because the rest of the spectrum is separated by large gaps, $\sim\delta_{\pm 2,0}$.

### C. (±1, $v_L$) sequence

**(-1,0).** For the $v$ =-1 QHFM state of primary Dirac fermions [equivalently, (-1,0) in Fig. 4], the superlattice broadening of the $N$=0 LL is expected to have a more drastic effect on spin polarization of electrons than in the case of the superlattice states discussed above. This is because the initial gap can be reduced by the formation of spin-valley textures (skyrmions) that appear around every electron added to the system. As a result, the broadening of the LL into a band not only reduces the gap,

$$\delta_{-1,0}(B < B_{1/1}) = E_{sk}(B) - D,$$

but can even close it if $D > E_{sk}(B)$. Note that only electrons in the occupied valley are engaged in the formation of skyrmions and reduce their energy, whereas states in the other valley are split by a large gap $E_C(B)$. Therefore, the latter states can be ignored as sketched in Fig. S8.

Because the collapse of the (-1,0) gap is clearly seen in Fig. 4 of the main text, we conclude that $E_C(B) > D > E_{sk}(B)$. The latter conclusion allows us to interpret the reentrance of the incompressible (-1,0) state at positive $B_{eff}$ (see Fig. 4c) by assigning it a gap

$$\delta_{-1,0}(B > B_{1/1}) = D - E_{sk}(B).$$

Indeed, upon the increase of $B$, the dense parts of the fractal spectrum shrink (Fig. S5). As shown in Fig. S8, this exposes a pair of LLs ($N'$=0 and $N''$=0), which are split by $D - E_{sk}(B) \ll E_{sk}(B)$. This explains the observed asymmetry in the gap behavior between $B<B_{1/1}$ and $B>B_{1/1}$.

**(-1,±1).** We further speculate that adding (subtracting) electrons would fill the $N''$=0 LL (deplete $N''$=0). For the fully occupied (emptied) states in this pair, which corresponds to $v_L$ = ($n_e$-$n_{v=-1}$)$\phi_0$/(B-$B_{1/1}$)= ±1, the incompressibility gap relies on a single-particle splitting between $N'$=0 and $|N'|$=1 LLs of third-generation Dirac fermions, $\sim v_*\sqrt{2\hbar e B_{eff}}$. The resulting gaps (see Fig. S8) are

$$\delta_{-1,\pm 1}(B > B_{1/1}) = v_*\sqrt{2\hbar e B_{eff}} - [D - E_{sk}(B)]$$

and have a small size as compared to the (-2, 1) gap. Moreover, the absence of the 0-th LLs at the edges of the central band at $B <B_{1/1}$ (see Fig. S5) suggests even smaller gaps between $|N'|$=1 and $|N'|$=2 in the much denser part of the spectrum for third-generation Dirac fermions, in agreement with experimental observations in Fig. 4.

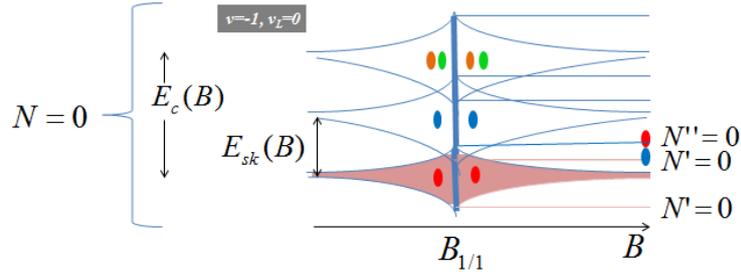

**Fig. S8. (±1, $\nu_L$) sequence of QHFM states: reverse Stoner transition and band collapse.** The QHFM state at $\nu$ =-1 is considered here. The occupied bands are again shown in pink but, for clarity, we avoid coloring of empty Hofstadter minibands that are only contoured in blue.

To conclude this section, the above considerations offer qualitative understanding of the entire hierarchy of gaps ($\nu, \nu_L$) observed experimentally. Largest are, as expected, the single-particle gaps (±2,0) that are of the order of the cyclotron energy for primary Dirac fermions in $B$ ~25T. The next in terms of the gap size is the incompressible state (0,0) with $\delta$ determined by the Coulomb energy with the magnetic length scale given by the total field $B$ ~$B_{1/1}$. Then, states (-2,1), (-2,2), (2,-1), (2,-2), (0,1) and (0,-1) follow. They exhibit the gaps determined by weaker exchange interaction at the 0-th LLs of third-generation Dirac fermions and are set by the Coulomb energy in $B_{eff}$ =$B$-$B_{1/1}$. Finally, two states (-1,0) and (1,0) undergo a reverse Stoner transition and exhibit a relatively large exchange gap at $B$ <$B_{1/1}$ (determined by skyrmion effects) with a much smaller gap at $B$ >$B_{1/1}$.